\documentclass[10pt]{article}
\usepackage{opex3}
 \usepackage{amsmath}
 \usepackage{amssymb}
 \usepackage{amsfonts}
  \usepackage{upgreek}
 
\def\U#1{{%
\def\O{\mbox{O}}
\def\u{\mbox{u}}
\mathcode`\u=\mu
\mathcode`\O=\Omega
\mathrm{#1}}}
\def\sub#1{_{\mbox{\scriptsize#1}}}
\def\sur#1{^{\mbox{\scriptsize#1}}}
\def\vct#1{{\mathchoice{\mbox{\boldmath$#1$}}{\mbox{\boldmath$#1$}}%
  {\mbox{\scriptsize\boldmath$#1$}}{\mbox{\scriptsize\boldmath$#1$}}}}
\def\jj{\,\mathrm{j}}                   

\begin{document}
\title{Freestanding transparent terahertz half-wave plate
using subwavelength cut-wire pairs}
\author{Yosuke Nakata,$^{1,*}$ Yudai Taira,$^{2}$ Toshihiro Nakanishi,$^{3}$ and Fumiaki Miyamaru$^{1,2}$}

\address{$^1$Center for Energy and Environmental Science, Shinshu University, 4-17-1 Wakasato, Nagano 380-8553, Japan\\
$^2$Department of Physics, Faculty of Science, Shinshu University, 3-1-1 Asahi, Matsumoto, Nagano 390-8621, Japan\\
$^3$Department of Electronic Science and Engineering, Kyoto University, Kyoto 615-8510, Japan}

\email{$^*$y\_nakata@shinshu-u.ac.jp} 



\renewenvironment{abstract}
{\vskip1pc\noindent\begin{center} \begin{minipage}{.8\textwidth} {\bf Abstract: \ } }
{ \\ \vskip-.5pc \noindent \small \copyright \, \number\year \hskip.05in
   Optical Society of America. One print or electronic copy may be made for personal use only. Systematic reproduction and distribution, duplication of any material in this paper for a fee or for commercial purposes, or modifications of the content of this paper are prohibited. \\ \hfil \end{minipage}\end{center}\normalsize\vskip-1.5pc}%

   \begin{abstract}
We designed a cut-wire-pair metasurface
that works as a transparent terahertz half-wave plate,
by matching the electric and magnetic resonances of the structure.
Due to the impedance matching nature of the resonances,
a large transmission phase shift between the orthogonal
polarizations was achieved, while permitting a high transmission.
The electric and magnetic responses of the proposed structure
were confirmed by evaluating the electric admittance and magnetic impedance.
The structure was fabricated on a flexible film
and its helicity-conversion
 function in the terahertz frequency range was experimentally demonstrated.
The thickness of the device is less than $1/10$ of the working vacuum wavelength, and
 a high amplitude helicity conversion rate over $80\, \%$ was achieved.
Finally, using simulations,
 we demonstrate the feasibility of the gradually rotating cut-wire-pair array in terahertz wave-front control.
   \end{abstract}


\bibliographystyle{osajnl}

 \section{Introduction}

During the last two decades, there has been rapid progress in terahertz spectroscopy techniques and they have been applied in several fields \cite{Lee2008}.
Especially, the recent progress in the development of high-power terahertz sources has accelerated the growth of terahertz nonlinear spectroscopy \cite{Kampfrath2013},
and imaging technologies \cite{Fukasawa2015}. In spite of these remarkable developments,
commercially available devices to control terahertz waves are still lacking.
The progress of terahertz polarization-dependent spectroscopy and imaging applications demand terahertz waveplates.

Although achromatic terahertz waveplates have been
developed \cite{Masson2006,Nagai2014}, the size of conventional terahertz devices is typically much larger than the terahertz wavelength.
To make the terahertz system compact,
it is highly desirable to develop
ultrathin waveplates.
Artificially designed
materials, or {\it metamaterials}, have been developed to overcome this problem \cite{Tao2011}.
Two-dimensional metamaterials called {\it metasurfaces}, which have thicknesses much less than working wavelengths, and
their ability to control electromagnetic waves
have been studied intensively \cite{Chen}.
So far, the application of terahertz metasurfaces in
quarter wave plates have been investigated \cite{He2013, Sieber2014, Wang2015a, Cong2014}.

A half-wave plate has the ability to convert helicity, e.g. from
right-circular polarization (RCP) to left-circular polarization (LCP).
However, in principle, a highly transparent half-wave plate with an amplitude
conversion efficiency from
LCP to RCP (RCP to LCP) over $50\,\U{\%}$ cannot be realized using a single-layer metasurface \cite{Ding2015}.
Therefore, multilayer designs are required to achieve this functionality.
Recently, Wu {\it et al.} discovered that two layers of double-cut split ring resonators work as a highly transparent subwavelength half-wave plate,
and experimentally realized them in the microwave frequency range \cite{Wu2016};
however, the guiding principle used to design the structure is unclear.
Another approach to realize a highly transparent half-wave plate is the use of a multilayer dipole array.
Y.~He {\it et al.} proposed the use of three-layer dipole structures to implement
an infrared half-wave plate, based on the microwave phase-shifter theory \cite{He2013}.
However, in order to make the fabrication process simpler.
it is desirable have a lower number of dipole layers.

In this paper, we show that a metasurface composed of
cut-wire pairs, interacting with each other,
can act as a highly transparent half-wave plate in the terahertz frequency range.
Theoretically, the metasurface is designed
by matching the electric and magnetic resonances of cut-wire pairs.
In spite of the extremely simple design of the metasurface,
it exhibited broadband impedance matching during simulations.
We fabricated the structure on a flexible film, and
demonstrated its helicity-conversion capability in the terahertz frequency range.
The application of a metasurface in terahertz wave-front control is also discussed.

 \section{Design\label{sec:2}}

\begin{figure}[t!]
\centering\includegraphics[width=7cm]{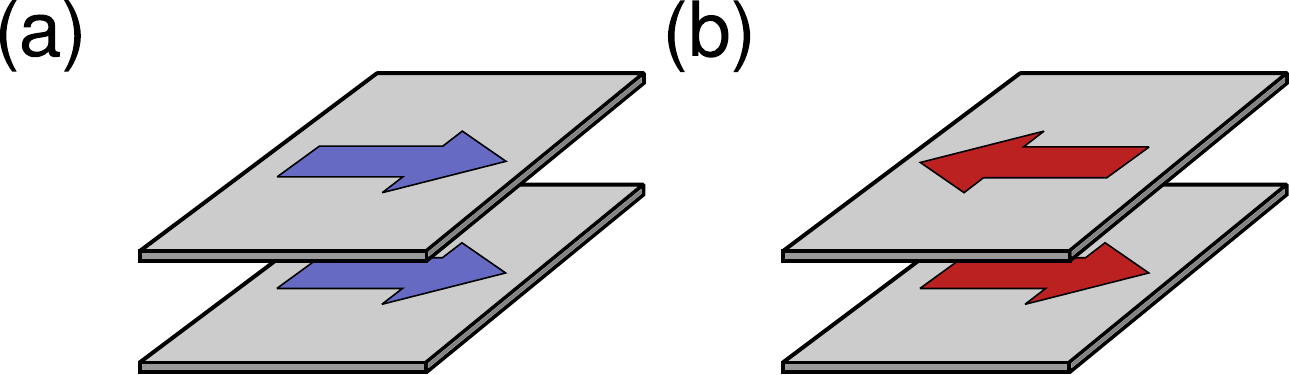}
\caption{\label{fig1}(a) Symmetric electric mode. (b) Anti-symmetric magnetic mode of the current on a metallic cut-wire pair.}
\end{figure}
From a macroscopic point of view, an isotropic metasurface can be
characterized by a zero-thickness surface with
an effective electric sheet admittance $Y\sub{e}$ and
and magnetic sheet impedance $Z\sub{m}$.
We assume that this surface is located at $z=0$.
The macroscopic parameters relate the effective surface electric current density $\vct{K}$
and the magnetic current density $\vct{K}\sub{m}$, flowing on $z=0$, with the coarse grained electric field $\vct{E}$ and magnetic field $\vct{H}$ at $z=0$, as
$\vct{K}=Y\sub{e}\vct{E}\sub{av},
\vct{K}\sub{m}=Z\sub{m}\vct{H}\sub{av},$
where ``av'' indicates that the fields on the upside ($z=+0$) and downside ($z=-0$) are averaged.
For a normally incident plane wave, the complex electric amplitude transmission and reflection coefficients are denoted as
 $t$ and $r$, respectively.
 The effective parameters $Y\sub{e}$ and $Z\sub{m}$ can be given by the following equations \cite{Pfeiffer2013}:
\begin{equation}
 \frac{Y\sub{e}}{Y_0}= 2\frac{1-r-t}{1+r+t},\hspace{2em}
\frac{Z\sub{m}}{Z_0}=2\frac{1+r-t}{1-r+t}.  \label{eq:1}
\end{equation}
Here, $Z_0=\sqrt{\mu_0/\varepsilon_0}(=1/Y_0)$ is the impedance of free space.
Defining $y\sub{e}=Y\sub{e}/(2Y_0)$ and $z\sub{m}=Z\sub{m}/(2Z_0)$,
we also have
$
 t=(1-y\sub{e}z\sub{m})/[(1+y\sub{e})(1+z\sub{m})],
 r=(z\sub{m}-y\sub{e})/[(1+y\sub{e})(1+z\sub{m})].
$
From these equations, the impedance matching $z\sub{m}=y\sub{e}$
leads to zero reflection, although the phase of $t$ could be an arbitrary value.
For example, $t=(1-\jj b)/(1+\jj b)$ induces the phase $\varphi=-2\arctan b$
for $y\sub{e}=z\sub{m}=\jj b$, $b\in \mathbb{R}$.

\begin{figure}[b!]
\centering\includegraphics[width=10.6cm]{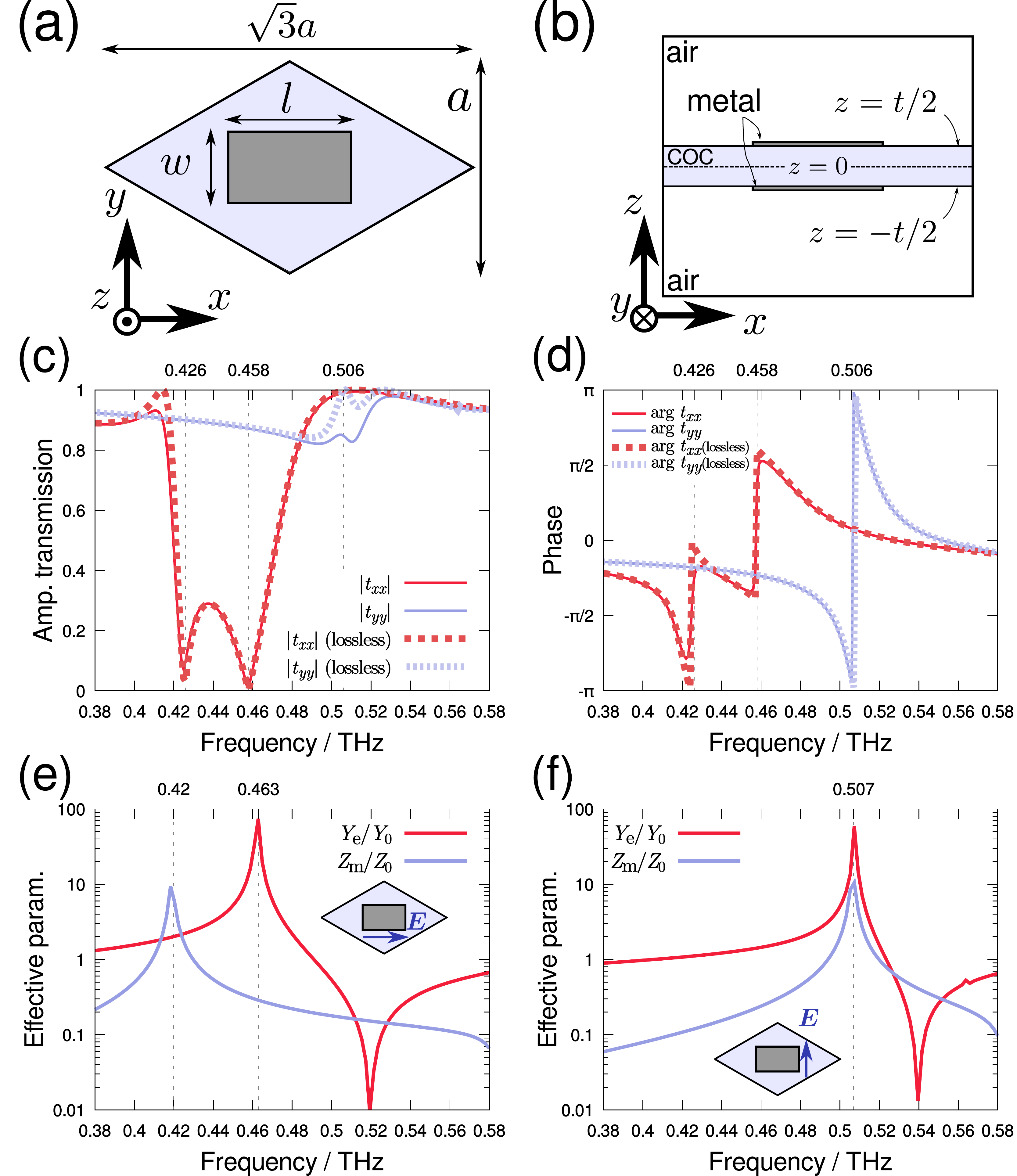}
\caption{\label{fig2} (a) Top view of the simulation model. (b) Side view of the model. (c) Amplitude and (d) phase transmission spectra of silver cut-wire pairs on a COC substrate with $a=585\,\U{\upmu m}$, $l=215\,\U{\upmu m}$, $w=169\,\U{\upmu m}$, and $t=40\,\U{\upmu m}$ for $x$ and $y$ polarized normally incident waves (lossless cases are also shown for comparison). Macroscopic electric admittance and magnetic impedance of the silver cut-wire pairs on a COC substrate for (e) $x$-polarized and (f) $y$-polarized waves.}
\end{figure}

Now, we consider a metasurface composed of metallic cut-wire pairs \cite{Dolling2005,Shalaev2005,Lam2007,Lam2008,Nguyen2011,Viet2015}.
A pair of parallel cut wires coupled with each other has two resonant modes:
an electric mode formed by the symmetrically flowing currents [Fig.~\ref{fig1}(a)],
and a magnetic mode formed by the antisymmetrically flowing currents [Fig.~\ref{fig1}(b)].
The electric mode can be induced by an oscillating electric field parallel to the current.
On the other hand,  a time-varying magnetic flux penetrating the cut-wire pair
will excite the magnetic mode composed of rotating currents in the cut-wire pair.
Thus, the electric and magnetic modes affect $Y\sub{e}$ and $Z\sub{m}$, respectively.
Matching the resonant frequencies of these modes could
induce a large phase shift, while permitting a high transmission.
Based on this idea, the structure was designed using the finite element method with COMSOL.
For this, we considered the cut-wire pairs to be arranged in the form of hexagonal lattices.
Note the following design strategy is applicable
to other lattice symmetries including square lattices.
The unit cell of the structure is depicted in Figs.~\ref{fig2}(a) and \ref{fig2}(b).
Periodic boundary conditions were utilized for the simulation.
The metallic cut wires were placed on both sides of
the substrate, which was made of cyclic olefin copolymers (COC) having a thickness $t=40\,\U{\upmu m}$
and refractive index $\tilde{n}=1.53-0.001\jj$ \cite{Wietzke2011}.

\begin{table}[!b]
\begin{center}
 \caption{\label{tab1} Calculated eigenfrequencies for cut-wire pairs on $40\,\U{\upmu m}$-thick  cyclo olefin copolymer substrate with $a=585\,\U{\upmu m}$, $l=215\,\U{\upmu m}$, and $w=169\,\U{\upmu m}$.}
 \begin{tabular}{c|c|c}
& electric mode & magnetic mode\\\hline
current $\parallel x$   & $0.473\,\U{THz}$ & $0.422\,\U{THz}$\\
current $\parallel y$   & $0.514\,\U{THz}$  &$0.510\,\U{THz}$
 \end{tabular}
\end{center}
\end{table}

First, the eigenfrequencies of the electric and magnetic modes were matched.
The eigenfrequencies of the metasurface were calculated by using the eigensolver function in COMSOL. In order to reduce the simulation complexity for the eigenfrequency analysis,
metallic cut wires were assumed to be zero-thick perfect electric conductor (PEC),
and only the $z\geq 0$ domain was considered.
For the magnetic modes, the PEC boundary condition was imposed on $z=0$.
For the electric modes, the perfectly magnetic conductor
boundary condition was imposed on $z=0$.
The perfectly matched layers were attached to the upper boundary of the air domain, so that there would be a proper absorption of the leakage radiation in the $z$ direction.
After optimization,
we obtained $a=585\,\U{\upmu m}$, $l=215\,\U{\upmu m}$, and $w=169\,\U{\upmu m}$.
The calculated eigenfrequencies of the structure are shown in Table~\ref{tab1}.
It can be seen that both the electric and magnetic modes of the current $\parallel y$
exist around $0.51\,\U{THz}$.
These modes may lead to the impedance matching of the structure, and could
induce a large phase shift while maintaining a high transmission.

Next, we calculated the transmission property of the metasurface.
A plane wave normally enters the metasurface from $z>0$ to $z<0$.
The metallic cut wire was treated as a transition boundary equivalent to
a $0.5\,\U{\upmu m}$-thick silver film having a conductivity of $46\,\U{S/\upmu m}$ \cite{Laman2008}.
For comparison, we also calculated transmission spectra of
a lossless metasurface composed of PEC cut-wire pairs
with a lossless substrate having a refractive index $\tilde{n}=1.53$.
For a $\beta$-polarized incident electric field $E_0\vct{e}_\beta\exp(\jj kz)$ in $z>0$, the
$\alpha$-polarized component of the transmitting electric field in $z<0$
is represented by $E_0\vct{e}_\alpha t_{\alpha \beta}\exp(\jj kz)$,
using amplitude transmission coefficients $t_{\alpha\beta}$,
a constant $E_0$,
the wavevector $k$ $(>0)$ in air,
and two orthonormal vectors $\vct{e}_\alpha$ $(\alpha=1,2)$
perpendicular to the $z$-axis.
The calculated amplitude and phase transmission spectra of the optimized
metasurface for $x$ and $y$ polarized plane waves are shown in Figs.~\ref{fig2}(c) and \ref{fig2}(d).
At $0.506\,\U{THz}$,
the functions of the half-wave plate can be observed to be
$|t_{xx}|\approx 0.99$, $|t_{yy}|\approx 0.85$,
and $\mathrm{arg}(t_{xx})-\mathrm{arg}(t_{yy})\approx 3.2$ for the silver metasurface.
Surprisingly, a high transmission amplitude $|t_{yy}|$
over 0.8 is observed in the broadband region.
This indicates that the oscillator strength and line width
of the electric and magnetic resonant modes are approximately balanced.
In comparison with the lossless metasurface, $|t_{yy}|$ of the silver metasurface
at the working frequency is degraded by the metal and substrate loss.
Cryogenic temperatures can be a path towards decreasing unavoidable loss effect \cite{Singh2010}.

The electric and magnetic responses of the silver metasurface for each polarization,
$|Y\sub{e}|$ and $|Z\sub{m}|$, calculated by Eq.~(\ref{eq:1}), are shown in Figs.~\ref{fig2}(e) and \ref{fig2}(f).
For an $x$-polarized wave, it can be seen that the frequencies of $|Y\sub{e}|$ and $|Z\sub{m}|$ peaks
are different, and they are far-off-resonant from the operation frequency
$0.506\,\U{THz}$.
Alternatively, for a $y$-polarized wave,
the peak frequencies are almost same, and impedance matching is approximately achieved in the broad band region.
Therefore, we can realize not only a half-wave plate,
but also an efficient phase plate with different phase shifts for different working frequencies.

 \section{Experimental demonstration of terahertz helicity conversion}

\begin{figure}[b!]
\centering\includegraphics[width=10cm]{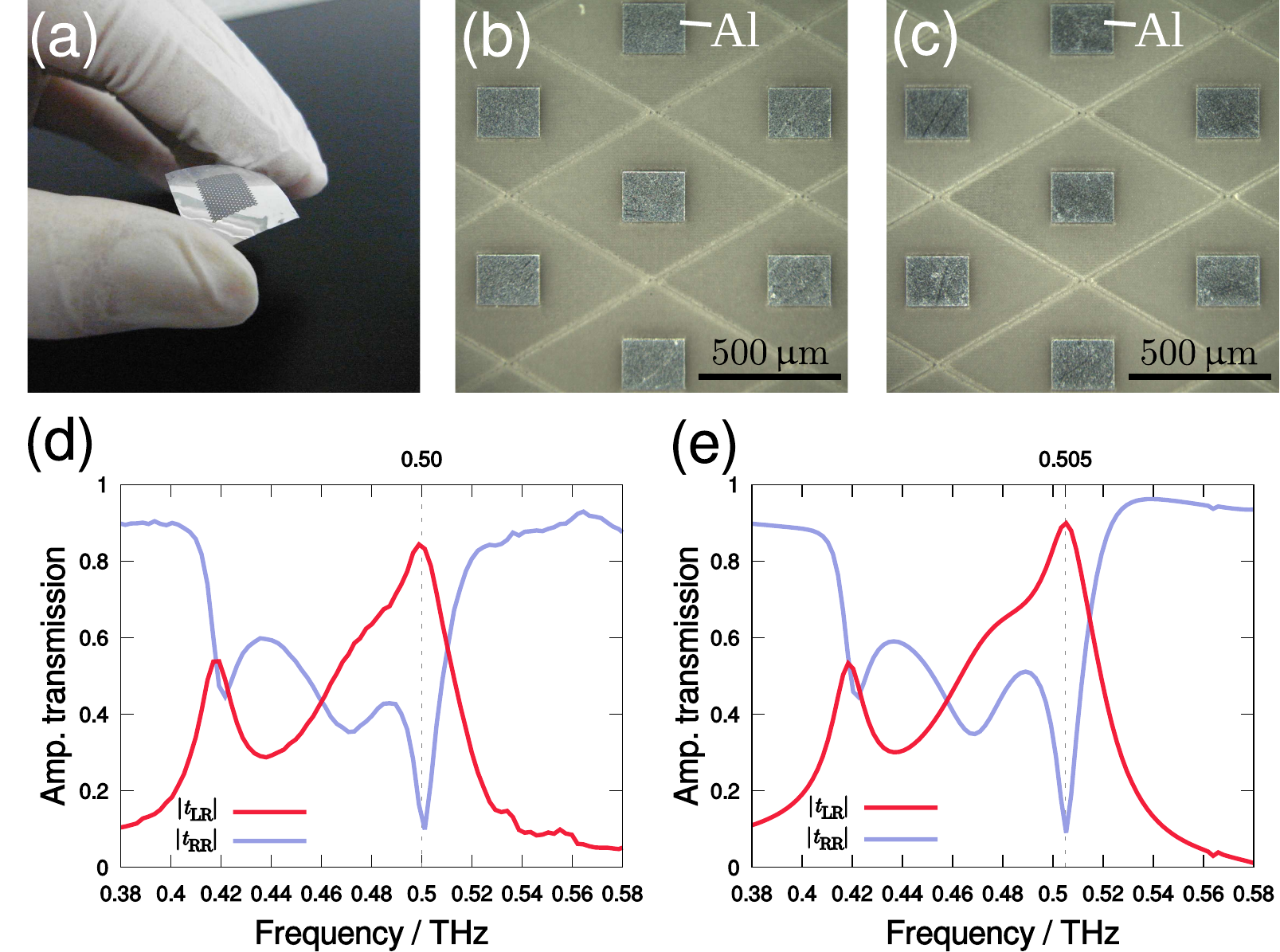}
\caption{\label{fig3} (a) Photograph of a sample. Microphotographs from (b) top and (c) bottom
 views of the sample. (d) Measured transmission spectra for the sample from the bottom side to the top side.
 (e) Simulated transmission spectra for $1.4\,\U{\upmu m}$-thick aluminum cut-wire pairs on a $40\,\U{\upmu m}$-thick COC substrate with $a=585\,\U{\upmu m}$, $l=217\,\U{\upmu m}$, and $w=174\,\U{\upmu m}$.}
\end{figure}

 The experimental demonstration of the function of the terahertz half-wave plate is described below.
 The sample was prepared as follows:
 (i) A $10\,\U{nm}$ titanium adhesion layer followed by
a $1.4\,\U{\upmu m}$ aluminum layer
 were deposited on both sides of the $40\,\U{\upmu m}$ COC (ZEONOR$^\text{\textregistered}$, Zeon Corp.) by electron-beam evaporation at room temperature, (ii) cut-wire-pair structures were formed on both sides of the substrate using a femtosecond laser ablation process.
Note that we deposited a $1.4\,\U{\upmu m}$-thick aluminum layer to avoid the skin effect as much as possible, but a
thickness sufficiently larger than the skin depth $\approx 150\,\U{nm}$ of aluminum at $0.5\,\U{THz}$ could be enough.
  The fabricated sample is shown in Figs.~\ref{fig3}(a)--(c).
In Figs.~\ref{fig3}(b) and \ref{fig3}(c), it can be seen that the boundary of the unit cell is whiter than the other parts. This is because the boundary is shaved twice by the scanning laser beam.
From the microphotographs of the sample, the
measured dimensions are $l=217\,\U{\upmu m}$ and $w=174\,\U{\upmu m}$, with a deviation of $1\,\U{\upmu m}$.

Next, using a terahertz time-domain spectroscopy system,
we demonstrated the helicity conversion function of the half-wave plate,
which is applicable to the wave-front control discussed in Sec.~\ref{sec:4}.
Here, the co- and cross-polarized amplitude transmissions $t_{xx}$ and $t_{yx}$ ($t_{yy}$ and $t_{xy}$) for normally incident
$x$ ($y$) polarized waves were measured using a standard cross ellipsometry polarization probing setup \cite{Morikawa2006},
and the values of $|t\sub{LR}|$ and $|t\sub{RR}|$ were calculated, where $\mathrm{R}$ and $\mathrm{L}$ represent RCP and LCP (rotating direction as seen from the detector side), respectively.
The plots of the experimentally obtained $|t\sub{LR}|$ and $|t\sub{RR}|$ are shown in Fig.~\ref{fig3}(d). In this figure, it can be observed that $|t\sub{LR}|$ peaks over 0.8, which is beyond the limit of conversion efficiency of a single-layer metasurface \cite{Ding2015},
at $0.50\,\U{THz}$. For comparison, the numerically calculated
$|t\sub{LR}|$ and $|t\sub{RR}|$ for cut-wire pairs of $1.4\,\U{\upmu m}$-thick aluminum having a conductivity $22\,\U{S/\upmu m}$ \cite{Laman2008}
with $a=585\,\U{\upmu m}$, $l=217\,\U{\upmu m}$, $w=174\,\U{\upmu m}$,
and $t=40\,\U{\upmu m}$, are shown in Fig.~\ref{fig3}(e), where the thin titanium layer was ignored in the simulation. Simulated maximum conversion efficiency is 0.9, which is
degraded by the loss effect discussed in Sec.~\ref{sec:2}.
The experimentally obtained spectra agrees well with the simulated spectra,
although the experimentally obtained peak value of helicity conversion is not as high as the simulated one. This can be attributed to fabrication error including over shaving of the substrate.
There is still room for device optimization; however, the basic function of the half-wave plate was demonstrated, and the thickness of the device is less than $1/10$ of the working vacuum wavelength.

 \section{Towards terahertz wave-front control\label{sec:4}}

In this section, we discuss the applicability of the cut-wire pairs to wave-front control.
If the structures that convert helicity are rotated, the converted wave attains the Pancharatnam--Berry phase.
Therefore, it is possible to control the wave front
by gradually rotating the structures \cite{Ding2015}.
 The Pancharatnam--Berry phase induced for
{\it reflected} circular-polarized waves is utilized for the highly efficient control of wave fronts
in the near-infrared region \cite{Zheng2015, Ding2015a}.
Recently, active control of the working frequency for reflected circular-polarized microwaves was theoretically proposed \cite{Xu2016}.
 The wave-front control for {\it transmitting} circular-polarized
 waves using three-layer metallic dipoles \cite{He2014}
 and dielectric resonators \cite{Zhao2016}
 has been studied with the help of simulations.

\begin{figure}[t!]
\centering\includegraphics[width=12cm]{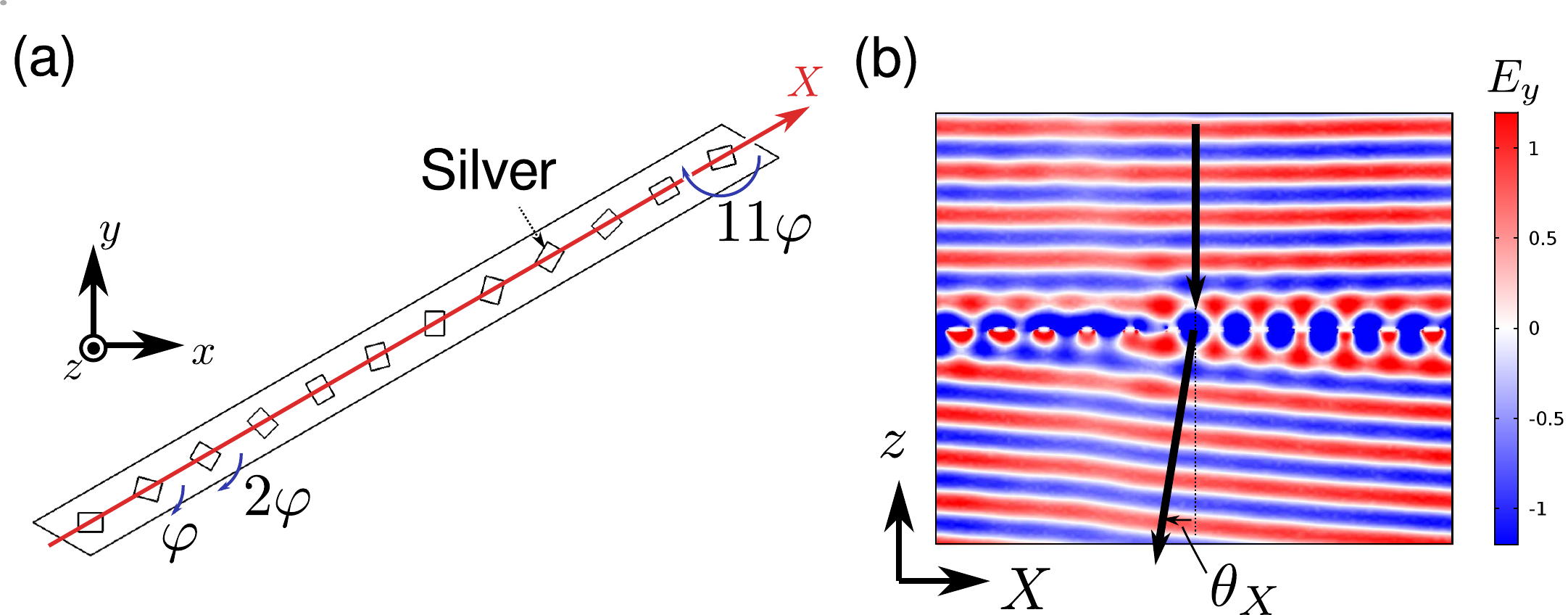}
\caption{\label{fig4} (a) The top view of simulation model. (b) $E_y$ at $0.504\,\U{THz}$ on $X$--$z$ plane.}
\end{figure}

To investigate the possibility of controlling a terahertz wave front,
we also performed a simulation using COMSOL for implementing a beam deflector.
For this, 12 unit cells [shown in Fig.~\ref{fig1}(a)] were connected and
each meta-atom was rotated, as shown in Fig.~\ref{fig4}(a).
The basic setup for the simulation was the same as that in Sec.~\ref{sec:2}.
The angle of rotation of each unit cell was increased by $\varphi$ from the previous adjacent unit cell,
with $\varphi=15^\circ$. The periodic boundary conditions were applied to
the side of the simulation domain.
From $z>0$, an RCP wave enters the $0.5\,\U{\upmu m}$-thick silver cut-wire array
on the surfaces ($z=-t/2,\ t/2$) of a COC substrate with a refractive index $\tilde{n}=1.53-0.001\jj$.
Perfectly matched layers were attached to the upper and bottom boundaries of the air domain for the proper absorption of diffracted waves.
The calculated $E_y$ at $0.504\,\U{THz}$ on the $X$--$z$ plane
is shown in Fig.~\ref{fig4}(b).
In Fig.~\ref{fig4}(b), we can see the beam is deflected by $\theta_X$ on the $X$--$z$ plane,
although the equiphase lines of the transmitted wave are not perfectly straight lines.
The observed $\theta_X\approx 5^\circ$
shows good agreement with the theoretically calculated value $\theta_X\sur{(theo)}\approx 4.9^\circ$.
The transmitting power evaluated by the integration of the Poynting vector is $73\,\%$.
In order to experimentally demonstrate this functionality in the terahertz frequency range, accurate fabrication and terahertz imaging to evaluate the sample are required.

 \section{Conclusion}

 In this study, we designed an ultrathin terahertz half-wave plate by matching the electric and magnetic resonances of cut wires coupled with each other.
The electric and magnetic responses were analyzed by evaluating the effective
electric admittance and magnetic impedance.
These results indicated that the electric and magnetic resonances
lead to a large phase shift while maintaining high broadband transmission.
The designed sample was fabricated using a femtosecond laser ablation process.
Next, we experimentally demonstrated its capability for the polarization conversion of a circularly polarized wave, with a frequency of approximately $0.5\,\U{THz}$.
The thickness of the device is less than $1/10$ of the working wavelength, and the observed amplitude helicity conversion efficiency is over $80\%$, which is
beyond the theoretical limit of $50\%$ that can be achieved by a single-layer metasurface.
Further optimization of the design and fabrication process could lead to higher conversion efficiency.
Although the working bandwidth of the metasurface is limited to the narrow band, its ultrathin size can be utilized for applications
involving continuous terahertz waves, which are often used in terahertz communication technology.
Finally, we discussed the possible application of the metasurface to highly efficient wave-front control. A beam deflector can be realized by gradually rotating the meta-atoms.
The experimental realization of such wave-front control could be studied in future works.

\section*{Acknowledgments}
The authors thank Yoshiro Urade for the useful discussions.
The electron-beam evaporation was performed with the help of Kyoto University Nano Technology Hub, as part of the ``Nanotechnology Platform Project'' sponsored by MEXT in Japan.

\end{document}